\documentstyle[12pt]{article}
\begin{document}

\def\beq{\begin{equation}}
\def\eeq{\end{equation}}
\def\bce{\begin{center}}
\def\ece{\end{center}}
\def\bea{\begin{eqnarray}}
\def\eea{\end{eqnarray}}
\def\ben{\begin{enumerate}}
\def\een{\end{enumerate}}
\def\ul{\underline}
\def\ni{\noindent}
\def\nn{\nonumber}
\def\bs{\bigskip}
\def\ms{\medskip}
\def\wt{\widetilde}
\def\wh{\widehat}
\def\brr{\begin{array}}
\def\err{\end{array}}

\hfill UB-ECM-PF 96/4

\hfill hep-th/9602113

\hfill February 1996

\vspace{5mm}

\bce
{\bf \Large Antisymmetric tensor fields on spheres: \\
functional determinants and non--local counterterms}
\ece
\vspace{3mm}
\bce
{\bf E. Elizalde}\footnote{E-mail: eli@zeta.ecm.ub.es}, 
{\bf M. Lygren}\footnote{E-mail: lygren@fm.unit.no}\\
Center for Advanced Studies CEAB, CSIC, Cam\'{\i} de Santa
B\`arbara,
17300 Blanes,
\\ and  Department ECM and IFAE, Faculty of Physics,
University of Barcelona, \\ Diagonal 647, 08028 Barcelona,
Spain \\
and \\
{\bf D.V. Vassilevich} \\
Department of Theoretical Physics, St. Petersburg University, \\
198904 St. Petersburg, Russia
\ece

\vspace{8mm}

\begin{abstract}
The Hodge--de Rham Laplacian on spheres acting on antisymmetric
tensor fields is considered. Explicit expressions for the spectrum
are derived in a quite direct way, confirming previous results. 
Associated functional determinants and the heat kernel expansion
are evaluated. Using this method, new non--local counterterms in the
quantum effective action are obtained, which can be
 expressed in terms of Betti numbers.
\end{abstract}

\vfill

\newpage

\bce {\large \bf I. Introduction} \ece

Modern interest in antisymmetric tensor fields is
connected with supergravity theories where these
fields appear as members of a supermultiplet. Kaluza--Klein
compactification of higher dimensional supergravities
leads to backgrounds of the form $S^d\times R^4$, where
$S^d$ is the $d$-dimensional sphere. Quantum effects on
such backgrounds were considered e.g. in \cite{CanW,FrT}
Some general mathematical
statements about $p$-forms ---very useful in this context---
can be found in the monography
\cite{PBG}. Typically, the
action for an antisymmetric tensor field $B$ reads
\begin{equation}
S=\int \sqrt g dx F_{ij...k} F^{ij...k}, \quad F=dB , \label{eq:act}
\end{equation}
where $d$ denotes external differentiation of forms.  The
action (\ref{eq:act}) for the $p$-form $B^p$ is invariant under gauge
transformations
\begin{equation}
B^p \to B^p+dB^{p-1}. \label{eq:gaug}
\end{equation}
All quantum corrections in a theory described by the action
(\ref{eq:act}), including the contribution of ghosts, can be
expressed in terms of determinants of the Hodge-de Rham
Laplacian
\begin{equation}
\Delta_{\rm HdR} =-(d^*d+dd^*). \label{eq:HdR}
\end{equation}
The spectrum of $\Delta_{\rm HdR}$ is the same for $p$- and
($d$-$p$)-forms. Consequently, it is enough to evaluate the
determinant of $\Delta_{\rm HdR}$ for $p\le d/2$. Due to the gauge
invariance under the transformation
(\ref{eq:gaug}) and to the factorization property
\begin{equation}
\det_p (-\Delta_{\rm HdR})=
\det_{pT} (-\Delta_{\rm HdR}) \times
\det_{(p-1)T} (-\Delta_{\rm HdR}) \label{eq:fa}
\end{equation}
---where the subscript $T$ means that the determinant is taken over the
space of transversal forms--- we can restrict our considerations to
transversal
$p$-forms. We will assume in all those expressions, as in (\ref{eq:fa}), 
that zero modes
(harmonic $p$-forms) are excluded from the determinants.

In Sect. 2 we will obtain the spectrum of the Hodge-de Rham
 Laplace operator  on the unit sphere
$S^d$, for any dimension $d$, acting on transversal $p$-forms.
In Sect. 3 we will calculate explicitly the determinants for the sphere,
and in Sect. 4 the heat kernel coefficients. A complete
list will be given up to
$d=7$, a dimension that is important in the compactification of
supersymmetric theories. Finally, a discussion on the transversal 
Laplacian and non-local counterterms will be provided in Sect. 5.
It is proven there that the heat kernel expansion for the
Hodge-de Rham Laplacian on transversal $p$-forms contains a constant term
even in the case of odd-dimensional spaces.
The new non--local counterterms in the
quantum effective action
will be expressed in terms of Betti numbers.
\ms

\bce {\large \bf II. Spectrum of the Laplace operator} \ece

    In this section we define    the spectrum of
the Laplace operator $\Delta_{HdR}$ on the unit sphere
$S^d$ acting on transversal $p$-forms. For $p=0,1$
this spectrum is well known \cite{RO}:
\begin{eqnarray}
b^0_l & =& -l(l+d-1), \quad  D_l^0  =
\frac {(2l+d-1)(l+d-2)!}{l!(d-1)!}, \quad
l  =0,1,2,\dots \nonumber \\
b^1_l & =& -l(l+d-1)+2-d, \quad D_l^1  =
\frac {l(l+d-1)(2l+d-1)(l+d-3)!}{(d-2)!(l+1)!},
\nonumber \\
 & \ & \ l  =1,2,3,\dots \label{eq:p01}
\end{eqnarray}
where $b_l^p$ denote the eigenvalues and $D_l^p$  their
degeneracies. For higher forms the spectrum of the Laplace
operator on $S^d$ can be obtained by using standard
group theoretical techniques \cite{IK}--\cite{Dplb}.

For any
homogeneous space $G/H$ a field $\Phi_A$ belonging to an
irreducible representation $D(H)$ can be expanded as \cite{SaSt}
\begin{equation}
\Phi_A (x)=V^{-\frac 12} \sum_{n, \zeta ,q}
\sqrt {\frac {d_n}{d_D}} D^{(n)}_{A\zeta ,q}
(g_x^{-1}) \phi^{(n)}_{q,\zeta }\ , \label{eq:(2.4)}
\end{equation}
where $V$ is the volume of $G/H$ and $d_D=$ dim $D(H)$.
We sum over representations $D^{(n)}$ of $G$
which give $D(H)$ after reduction to $H$. Here $\zeta$ labels
the multiple components  $D(H)$ in the branching
$D^{(n)}\downarrow H$, $d_n=$ dim $D^{(n)}$. The matrix
elements of $D^{(n)}$ have the following orthogonality
property
\begin{equation}
\int_{G/H} dx \sqrt g
D^{(n) *}_{A\zeta ,q} (g_x^{-1})
D^{(n')}_{A\xi ,p} (g_x^{-1})=
Vd_n^{-1} d_D \delta_{\zeta \xi} \delta_{p q}
\delta_{nn'}. \label{eq:(2.5)}
\end{equation}

Consider, for example, the case when $d=5$: $S^5=SO(6)/SO(5)$.
The representations $D(H)=D(SO(5))$ describing antisymmetric
tensors are just antisymmetric tensor powers of the vector
representation of $SO(5)$:
\begin{eqnarray}
p=0 &\quad & D(SO(5))=[0,0] \nonumber \\
p=1 &\quad & D(SO(5))=[1,0] \nonumber \\
p=2 &\quad & D(SO(5))=[1,1] \nonumber \\
p=3 &\quad & D(SO(5))=[1,1] \label{eq:doth}
\end{eqnarray}
We label the representations by their Dynkin indices in square
brackets. Notice that we use here the slightly non-standard definition
of the Dynkin indices from the book \cite{BaRon}. These indices
are more convenient for the reduction of representations.
Owing to duality, the representations in the last two
lines of (\ref{eq:doth}) are equivalent.

In order to construct the harmonic expansion (\ref{eq:(2.4)}) one must
find all the representations of $SO(6)$ which give the representations
(\ref{eq:doth}) after reduction to $SO(5)$. For a given
irreducible representation $[q_1,q_2]$ of $SO(5)$ with integer
Dynkin indices $q_1$ and $q_2$, the representations $[m_1,m_2,m_3]$
of $SO(6)$ containing $[q_1,q_2]$ are defined by the conditions
\cite{BaRon}
\begin{equation}
m_1 \ge q_1 \ge m_2 \ge q_2 \ge \vert m_3 \vert , \label{eq:noneq}
\end{equation}
where all $m_A$ are integers and $m_1$ and $m_2$ are non-negative.
Any representation satisfying the nonequality (\ref{eq:noneq})
contains the single representation $[q_1,q_2]$. Summation over
$\zeta$ in (\ref{eq:(2.4)}) may be omitted. One can easily
find all the representations of $SO(6)$ that are needed
\begin{eqnarray}
p=0 \quad & D^{(l)}(SO(6))& =[l,0,0],\ l=0,1,\dots \nonumber \\
p=1 \quad & D^{(l)}(SO(6))& =[l,1,0],\ l=1,2,\dots \nonumber \\
 \quad &               & =[l,0,0],\ l=1,2,\dots \nonumber \\
p=2,3 \quad & D^{(l)}(SO(6))& =[l,1,0],\ l=1,2,\dots \nonumber \\
 \quad &               & =[l,1,1],\ l=1,2,\dots \nonumber \\
 \quad &               & =[l,1,-1],\ l=1,2,\dots \label{eq:dotg}
\end{eqnarray}
External differentiation maps transversal $p-1$ forms to
longitudinal $p$-forms. This mapping can be traced back to the
corresponding
spherical harmonics. Hence, for the transversal forms only the following
representations contribute to the harmonic expansion:
\begin{eqnarray}
p=0 \quad & D^{(l)}(SO(6))& =[l,0,0], \nonumber \\
p=1 \quad & D^{(l)}(SO(6))& =[l,1,0], \nonumber \\
p=2 \quad & D^{(l)}(SO(6))& =[l,1,1], \nonumber \\
 \quad &               & =[l,1,-1], \nonumber \\
p=3 \quad & D^{(l)}(SO(6))& =[l,1,0],  \qquad l=1,2,3, \ldots
\label{eq:dtra}
\end{eqnarray}
 The scalar mode with $l=0$ belongs to the kernel
of the Laplace operator and should be regarded as an harmonic zero-form.

In the space of $p$-forms there are two main second-order
differential operators, namely the Hodge--de Rham Laplacian,
$-\Delta_{\rm HdR}=dd^*+d^*d$, and the ordinary Laplacian,
$\Delta =\nabla^i \nabla_i$.
On the sphere $S^d$ these two operators
differ by a constant, namely
\begin{equation}
-\Delta_{\rm HdR}=\Delta +p^2-dp .
\label{eq:laps}
\end{equation}

In any homogeneous space $G/H$ the Laplace operators can be
expressed in terms of the quadratic Casimir operators of $G$
and $H$:
\begin{equation}
\Delta =C_2(G)-C_2(H) ,\quad -\Delta_{\rm HdR}=C_2(G).
\label{eq:Cas}
\end{equation}
Using the harmonic expansion (\ref{eq:dtra}), (\ref{eq:doth}),
and standard expressions \cite{BaRon} for the Casimir operators
in (\ref{eq:Cas}), we obtain the eigenvalues $b^p_l$ of the Laplace
operator $\Delta_{HdR}$ acting on transversal $p$-forms on $S^5$. The
corresponding degeneracies $D^p_l$ are equal to the dimensions of the
representations of $SO(6)$.
\begin{eqnarray}
b_l^2 &=& -l(l+4)-4, \quad D_l^2=\frac 12 l(l+1)(l+3)(l+4),
\quad l=1,2,... \nonumber \\
b_l^3 &=& -l(l+4)-3, \quad D_l^3=\frac 13 l(l+2)^2(l+4),
\quad l=1,2,... \label{eq:d5p23}
\end{eqnarray}
For $p=0,1$ our result coincides with (\ref{eq:p01}).
Other spheres can be dealt with along the same lines.
For $d=3$ we have:
\begin{equation}
b_l^2=-l(l+2), \quad D_l^2=(l+1)^2, \quad l=1,2,...
\label{eq:d3p2}
\end{equation}
For $d=4$:
\begin{equation}
b_l^2=-l(l+3)-2, \quad D_l^2=\frac 12 l(2l+3)(l+3),
\quad l=1,2,... \label{eq:d4p2}
\end{equation}
For $d=6$:
\begin{equation}
b_l^2=b_l^3=-l(l+5)-6, \quad
D_l^2=D_l^3=\frac 1{12} l(l+1)(l+4)(l+5)(2l+5),
 \quad l=1,2,3,...
\label{eq:d6p23}
\end{equation}
For $d=7$:
\begin{eqnarray}
b_l^2=-l(l+6)-8,&\quad &
D_l^2=\frac 1{24} l(l+1)(l+3)^2(l+5)(l+6),\nonumber \\
b_l^3=-l(l+6)-9,&\quad &
D_l^3=\frac 1{18} l(l+1)(l+2)(l+4)(l+5)(l+6),
 \quad l=1,2,3,...
\label{eq:d7p23}
\end{eqnarray}
Owing to duality, the spectrum of
$\Delta_{\rm HdR}$ on transversal 4-forms on $S^7$ is the same as
on longitudinal 3-forms. The latter one coincides with that
on transversal 2-forms. Continuing in this way, one can define
the spectrum for higher values of $p$.
The remarkable property of equivalence of the spectra for
$p=2$ and $p=3$ on $S^6$ holds only for transversal
forms. In the above equations we have listed some spectra
for $p > d/2$. They are useful in some applications
not considered in this paper. For example, they are needed
for the computation of the spectrum of the Laplacian on a ball
\cite{Djmp,Dplb,InPr}.

There is a general function for the eigenvalues and their multiplicities,
which have been obtained above for some particular cases.
It is the following:
\bea
D_l(d,p) &=& \frac{(2l+d-1) \ (l+d-1)!}{p! \, (d-p-1)! \, (l-1)! \,
(l+p) (l+d-p-1)},  \nn \\
b_l (d,p) &=& -l(l+d-1)-p(d-p-1).   \label{gf1}
\eea
These equations can be obtained by means of lengthy but straightforward
calculations repeating step by step the above derivation of the
spectrum on $S^5$. All group theoretical techniques that are needed
can be learned from Chapts. 9 and 10 of Ref. \cite{BaRon}. Note that an
explicit derivation for the case of the ordinary Laplacian
had been carried out, e.g., in Ref. \cite{IK}, and that previous
results already existed in the
mathematical literature \cite{BeMi,IkTa}. Our method
is similar to the one in the papers \cite{BeMi,IkTa}.
However, explicit expressions for the eigenvalues and
degeneracies can be found in Ref. \cite{IK} only,
where reduction of the harmonic polynomials from $R^{d+1}$
was used. It is noticeable that a mistake in previous calculations
was reported in Ref. \cite{IK}, what shows that the computation is not
trivial at all. It thus seems useful
to present an alternative derivation of the spectrum,
which turns out to be in complete agreement with \cite{IK}.
Note that the eigenvalues of the Laplace operator on
transversal $p$-forms are denoted in Ref. \cite{IK}
by $\ ^{p+1}\lambda_{k}$, where $k=l-1=0,1,\dots$.
\ms

\bce {\large \bf III. Calculation of the determinants for the sphere}  \ece

Here we are going to calculate the determinants corresponding to the
Hodge-de Rham Laplacian on
spheres of different dimensions, $d=2,3,4,5,6,7$, and for forms of
different orders $p=1,2,3,4$. We shall make
 use the formulas (see above)
\beq
\det (-\Delta_{HdR})^{(d)} =
\det (-\Delta_{HdR})^{(d)}_{p^T}  \times
\det (-\Delta_{HdR})^{(d)}_{(p-1)^T}, \label{a1.1}
\eeq
and employ the definition of determinant through the zeta function of
the corresponding operator, that is
\beq
\det A = \exp \left( - {\zeta_A}' (0) \right).
\label{a1.2}
\eeq
Owing to the multiplicative property of the determinant ---which is
obviously fulfilled for the operators we are going to consider (see
\cite{eet1} for a discussion of more general cases)--- at the level of the
zeta functions the product in (\ref{a1.1}) transforms into a sum of the
corresponding zeta functions (even before taking the derivative). We
shall arrange our calculations according to this observation.
The general methods employed in \cite{eli1,eli2} will be used
(see \cite{eli3}, for more references to these techniques).

Using the general formulas (\ref{gf1}) for the spectrum and its
degeneracy, one can write the expression of the zeta function
corresponding to a $p$-form in any dimension $d$ ($p\leq (d+1)/2$),
namely
\bea
&&\zeta_{-\Delta^{(d)}_{p^T}} (s) \ = \ \sum_{l=1}^\infty D_l (d,p) \,
   [-b_l(d,p)]^{-s}  \\
&& \ = \frac{1}{p! \, (d-p-1)!} \sum_{l=1}^\infty \frac{(2l+d-1)(l+d-1)!}{(l-1)!
\, (l+p)(l+d-p-1)}
\left[ \left(l+\frac{d-1}2\right)^2 - \left( p- \frac{d-1}{2} \right)^2 \right]^{-s}
. \nn
\eea
To continue, we notice that the degeneracy is a polynomial in $l$ of order
$d-1$, and we expand it in powers of $l+(d-1)/2$:
\beq
 D_l(p,d)=\sum_{\alpha=0}^{d-1}e_{\alpha}(d,p)\left(l+\frac{d-1}{2}
\right)^{\alpha}.
\eeq
Formally, we can write
\beq
e_{\alpha}= \frac{1}{\alpha} \frac{d^\alpha}{dl^\alpha} \left. D_l(p,d)\right|_{l=\frac{1-d}2}.
\eeq
The sum over $l$ can be evaluated easily, e.g.
\bea
\zeta_{-\Delta^{(d)}_{p^T}} (s) & =& \sum_{\alpha=0}^{d-1}e_{\alpha}(d,p)
 \sum_{l=1}^{\infty}\left(l+\frac{d-1}{2}\right)^\alpha
    \left[\left(l+\frac{d-1}{2}\right)^2-
\left(p-\frac{d-1}{2}\right)^2 \right]^{-s} \nn \\
&=&\sum_{\alpha=0}^{d-1}e_{\alpha}(d,p) \sum_{l=1}^{\infty}\left(l
       +\frac{d-1}{2}\right)^{\alpha-2s}
      \left[1-\frac{(p-\frac{d-1}{2})^2}{(l+\frac{d-1}{2})^2} 
\right]^{-s} \label{gen.zeta} \\
&=&\frac{1}{\Gamma (s)} \sum_{\alpha=0}^{d-1}e_{\alpha}(d,p)
    \sum_{k=0}^\infty \frac{\Gamma (k+s)}{k!} \left(p-\frac{d-1}{2}
    \right)^{2k} \zeta_H (2s+2k-\alpha,\frac{d+1}{2}), \nn
\eea
where we have used the binomial expansion. Note that this expansion is
absolutely convergent,
since $[p-(d-1)/2]^2/[l+(d-1)/2]^2 < 1$ for $p\leq (d+1)/2$. 
The indetermined number $0^0$ when $p=(d-1)/2$ is consistently
defined to be one.
Here $\zeta_H(s,\nu)$ is the Hurwitz zeta-function. For $\nu$  a natural
 number,
this zeta-function can be directly related to the Riemann zeta-function
through the formula \cite{eli1}--\cite{eli3}
\beq
  \zeta_H (s,m) = \zeta_R(s)-\sum_{l=1}^{m}l^{-s}.
\eeq
For $\nu$  a half-integer, we can correspondingly subtract
terms from $\zeta_H (s,1/2)$, which is again  related to the Riemann
zeta-function
\beq
    \zeta_H (s,1/2) = (2^s-1)\zeta_R(s).
\eeq
For $d$ even, $e_\alpha=0$ for $\alpha=0,2,...,d-2$, and for $d$ odd,
$e_\alpha=0$ for $\alpha=1,3,...,d-2$. When we differentiate the
zeta-function (\ref{gen.zeta}) we must distinguish between these two
cases. For $d$ odd, we get
\bea
 &&\zeta'_{-\Delta^{(d)}_{p^T}} (0)=\sum_{a=0}^{\frac{d-1}{2}} e_{2a} \left[
    2 \zeta'_H (-2a,\frac{d+1}{2})+ \sum_{k=1}^{\infty} \frac{ \left(p-
          \frac{d-1}{2}\right)^{2k}}{k} \zeta_H (2k-2a,\frac{d+1}{2})
 \right].
\eea
While for $d$ even the expression is a bit more complicated since the
Hurwitz zeta-function has a pole when its argument is one. Using
the Laurent series expansion
\beq
 \zeta_H (2s+1,\nu) = \frac{1}{2s}-\Psi(\nu)+{\cal O} (s),
\eeq
we obtain, for $d$ even,
\bea
 \zeta'_{-\Delta^{(d)}_{p^T}} (0)&=&\sum_{a=0}^{\frac{d}{2}-1} e_{2a+1}
    \left[
    2 \zeta'_H (-2a-1,\frac{d+1}{2})+ \sum_{k=1}^{a} \frac{ \left(p-
       \frac{d-1}{2}\right)^{2k}}{k} \zeta_H (2k-2a-1,\frac{d+1}{2})
    \right. \nn \\
&+& \frac{ \left(p-\frac{d-1}{2}\right)^{2a+2}}{a+1}\left(
       \frac{1}{2} \sum_{l=1}^{a} l^{-1} -\Psi (\frac{d+1}{2}) \right) \nn\\&+&  \left. \sum_{a+2}^{\infty} \frac{ \left(p-
          \frac{d-1}{2}\right)^{2k}}{k} \zeta_H (2k-2a-1,\frac{d+1}{2})
         \right].
\eea

Continuing in this way and substituting for the derivatives of the
Riemann zeta
function the values \cite{eemc}
\bea
&& \zeta'(0) = -\frac{1}{2} \ln (2\pi), \qquad
\zeta'(-1)= -0.1654211437, \qquad
\zeta'(-2)= -0.0304484571, \nn \\ &&
\zeta'(-3)= 0.0053785764, \qquad
\zeta'(-4)= 0.0079838115, \qquad
\zeta'(-5)= -0.0005729860, \nn \\ &&
\zeta'(-6)= -0.0058997591, \ \ldots,
\label{a1.12}
\eea
we have obtained the following numerical results for the determinants:
\bea
\det (-\Delta_{HdR})^{(7)}_{4} &=&  0.088786, \nn \\
\det (-\Delta_{HdR})^{(7)}_{3} &=&  0.088786, \nn \\
\det (-\Delta_{HdR})^{(7)}_{2} &=&  1.858601, \nn \\
\det (-\Delta_{HdR})^{(7)}_{1} &=&  0.775194, \nn \\
\det (-\Delta_{HdR})^{(6)}_{3} &=&  7.103758, \nn \\
\det (-\Delta_{HdR})^{(6)}_{2} &=&  1.726306, \nn \\
\det (-\Delta_{HdR})^{(6)}_{1} &=&  0.835544, \nn \\
\det (-\Delta_{HdR})^{(5)}_{3} &=& 11.090330, \nn \\
\det (-\Delta_{HdR})^{(5)}_{2} &=& 11.090330, \nn \\
\det (-\Delta_{HdR})^{(5)}_{1} &=&  0.581303, \nn \\
\det (-\Delta_{HdR})^{(4)}_{2} &=&  0.128002, \nn \\
\det (-\Delta_{HdR})^{(4)}_{1} &=&  0.621433,
\label{a1.13} \\
\det (-\Delta_{HdR})^{(3)}_{2} &=&  0.095528, \nn \\
\det (-\Delta_{HdR})^{(3)}_{1} &=&  0.095528, \nn \\
\det (-\Delta_{HdR})^{(2)}_{1} &=&  10.210016. \nn
\eea
\ms

\bce {\large \bf IV. Calculation of the heat kernel coefficients} \ece

The heat-kernel coefficients $B_k$ are given from the small$-t$ expansion of the
heat kernel $K(t)$,
\beq
K(t)= (4\pi t)^{-\frac d 2} \sum_{k=0,\frac{1}{2},1,...} B_k t^k
\eeq
When we consider a manifold without boundaries, as is the case for the sphere,
the coefficients with a half-integer $k$ vanish. There is a close connection
between the coefficients for an operator and its zeta function. This
connection is given by the formulas:
\beq
\mbox{Res}\, \zeta (s)= \frac{B_{\frac{d}{2}-s}}{(4 \pi)^{\frac d 2}
 \Gamma (s)},
\label{eq:41}
\eeq
at $s=\frac m 2, \frac{m-1} 2 ,....,\frac 1 2 ; -\frac{2l+1} 2$ for $l=0,1,2..$
, and
\beq
\zeta(-m)= (-1)^m m!  \frac{B_{\frac{d}{2}+m}}{(4 \pi)^{\frac d 2}},
\label{eq:42}
\eeq
for $m=0,1,2,...$ These formulas constitute a very powerful approach to the
determination of the heat kernel coefficients (see, for instance, Refs.
\cite{bek,bgke}, and the references therein).
Again we consider separately the cases
$d$ odd and $d$ even. When $d$ is odd we only have to calculate
 the residues, since
there are only integer coefficients. The residues matching integer
coefficients are located at $s= d/2-m$, $m=0,1,2...$. When $m \leq (d-1)/2$
\beq
  \mbox{Res}\, \zeta_{-\Delta^{(d)}_{p^T}} \left(\frac d 2 -m\right)
       = \frac{1}2 \sum_{a=\frac{d-1}2-m}^{\frac{d-1}2}
       e_{2a} \frac{\Gamma (\frac{1}2+a)}{ \Gamma (\frac{d}2-m)
       (\frac{1-d}2+m+a)!}\left(p-\frac{d-1}{2}\right)^{1-d+2m+2a},
\eeq
and when $m> (d-1)/2$
\beq
  \mbox{Res}\, \zeta_{-\Delta^{(d)}_{p^T}} \left(\frac d 2 -m\right)= \frac{1}2 \sum_{a=0}^{\frac{d-1}2}
      e_{2a} \frac{\Gamma (\frac{1}2+a)}{ \Gamma (\frac{d}2-m)
       (\frac{1-d}2+m+a)!} \left(p-\frac{d-1}{2}\right)^{1-d+2m+2a}.
\eeq
For even dimension we must consider the point values at $s=-m$ and the
residues at $s= d/2-l$, $l=0,1,...,d/2-1$.
\bea
\zeta_{-\Delta^{(d)}_{p^T}} (-m)&=&\sum_{a=0}^{\frac{d}2-1} e_{2a+1} \left[ \sum_{k=0}^{m} \frac{(-1)^k
         m!}{(m-k)!k!} \left(p-\frac{d-1}{2}\right)^{2k} \zeta_H (2k-2a-1-2m,
         \frac{d+1}{2}) \right.  \nn \\
&& + \left.\frac{(-1)^m m! a!}{2(1+a+m)!} \left(p-\frac{d-1}{2}\right)^{2+
          2a+2m}   \right],
\eea
and we get
\beq
\mbox{Res}\, \zeta_{-\Delta^{(d)}_{p^T}} (\frac{d}2-l)= \frac 1 2 \sum_{a=\frac{d}2-1-l}^{\frac{d}2-1}e_{2a+1}
           \frac{a!}{\Gamma (\frac{d}2-l) (1-\frac{d}2+l+a)!}
          \left(p-\frac{d-1}{2}\right)^{2-d+2a+2l}.
\eeq
As the actual zeta functions are  sums of the zeta functions for
the transversal field, the heat kernel coefficients are also just sums
of the corresponding coefficients for the transversal fields. Using the two
equations (\ref{eq:41}) and (\ref{eq:42}) we immediately obtain the
coefficients from the formulas given above.

For $d=7$, $p=4$ and $p=3$, we have:
\bea
  B_0 &=& \frac{35\,\pi^4}{3} , \nn \\
  B_1 &=& {{-175\,{{\pi }^4}}\over 3}  , \nn \\
  B_2 &=&{{2009\,{{\pi }^4}}\over {18}} , \nn \\
  B_3 &=& {{-553\,{{\pi }^4}}\over 6}, \nn \\
  B_4 &=& {{159\,{{\pi }^4}}\over 8}  , \nn \\
  B_5 &=&{{167\,{{\pi }^4}}\over {24}} , \nn \\
  B_6 &=&{{1289\,{{\pi }^4}}\over {720}}   , \nn \\
  B_7 &=& {{613\,{{\pi }^4}}\over {1680}} , \nn \\
  B_8 &=&{{71\,{{\pi }^4}}\over {1152}}  , \nn \\
  B_9 &=&{{461\,{{\pi }^4}}\over {51840}}  , \nn \\
  B_{10} &=&{{271\,{{\pi }^4}}\over {241920}}  . \nn
\eea
For $d=7$, $p=2$:
\bea
  B_0 &=&7\,{{\pi }^4} , \nn \\
  B_1 &=&-21\,{{\pi }^4} , \nn \\
  B_2 &=& {{133\,{{\pi }^4}}\over {10}} , \nn \\
  B_3 &=&{{371\,{{\pi }^4}}\over {30}} , \nn \\
  B_4 &=&{{-229\,{{\pi }^4}}\over {40}}  , \nn \\
  B_5 &=&{{-1213\,{{\pi }^4}}\over {120}} , \nn \\
  B_6 &=&{{-2807\,{{\pi }^4}}\over {720}} , \nn \\
  B_7 &=& {{6483\,{{\pi }^4}}\over {2800}} , \nn \\
  B_8 &=&{{132847\,{{\pi }^4}}\over {28800}} , \nn \\
  B_9 &=& {{1050881\,{{\pi }^4}}\over {259200}} , \nn \\
  B_{10} &=&{{3147083\,{{\pi }^4}}\over {1209600}} ., \nn
\eea
For $d=7$, $p=1$:
\bea
  B_0 &=&{{7\,{{\pi }^4}}\over 3} , \nn \\
  B_1 &=&{{7\,{{\pi }^4}}\over 3} , \nn \\
  B_2 &=&{{-301\,{{\pi }^4}}\over {90}} , \nn \\
  B_3 &=&{{-203\,{{\pi }^4}}\over {30}} , \nn \\
  B_4 &=&{{-43\,{{\pi }^4}}\over {40}}  , \nn \\
  B_5 &=&{{949\,{{\pi }^4}}\over {120}} , \nn \\
  B_6 &=& {{6839\,{{\pi }^4}}\over {720}} , \nn \\
  B_7 &=&{{9823\,{{\pi }^4}}\over {8400}} , \nn \\
  B_8 &=&{{-282271\,{{\pi }^4}}\over {28800}} , \nn \\
  B_9 &=&{{-3734393\,{{\pi }^4}}\over {259200}} , \nn \\
  B_{10} &=&{{-3734393\,{{\pi }^4}}\over {403200}} . \nn
\eea
For $d=6$, $p=3$:
\bea
  B_0 &=&{{64\,{{\pi }^3}}\over 3} , \nn \\
  B_1 &=&{{-256\,{{\pi }^3}}\over 3} , \nn \\
  B_2 &=&128\,{{\pi }^3} , \nn \\
  B_3 &=&{{-75008\,{{\pi }^3}}\over {945}} , \nn \\
  B_4 &=&{{1472\,{{\pi }^3}}\over {135}}  , \nn \\
  B_5 &=&{{256\,{{\pi }^3}}\over {99}} , \nn \\
  B_6 &=&{{373376\,{{\pi }^3}}\over {405405}} , \nn \\
  B_7 &=&{{40448\,{{\pi }^3}}\over {81081}} , \nn \\
  B_8 &=&{{1373248\,{{\pi }^3}}\over {3828825}} , \nn \\
  B_9 &=&{{167651072\,{{\pi }^3}}\over {535687425}} , \nn \\
  B_{10} &=&{{65263383424\,{{\pi }^3}}\over {206239658625}}
 . \nn
\eea
For $d=6$, $p=2$:
\bea
  B_0 &=&16\,{{\pi }^3} , \nn \\
  B_1 &=&-48\,{{\pi }^3} , \nn \\
  B_2 &=&{{128\,{{\pi }^3}}\over 3} , \nn \\
  B_3 &=&{{176\,{{\pi }^3}}\over {315}} , \nn \\
  B_4 &=&{{-48\,{{\pi }^3}}\over 5}  , \nn \\
  B_5 &=&{{-1360\,{{\pi }^3}}\over {297}} , \nn \\
  B_6 &=& {{-171328\,{{\pi }^3}}\over {405405}} , \nn \\
  B_7 &=&{{80096\,{{\pi }^3}}\over {81081}} , \nn \\
  B_8 &=&{{37236464\,{{\pi }^3}}\over {34459425}} , \nn \\
  B_9 &=&{{52062832\,{{\pi }^3}}\over {59520825}} , \nn \\
  B_{10} &=&{{2616564224\,{{\pi }^3}}\over {3618239625}} . \nn
\eea
For $d=6$, $p=1$:
\bea
  B_0 &=&{{32\,{{\pi }^3}}\over 5} , \nn \\
  B_1 &=&0 , \nn \\
  B_2 &=&{{-128\,{{\pi }^3}}\over {15}} , \nn \\
  B_3 &=&{{-1408\,{{\pi }^3}}\over {315}} , \nn \\
  B_4 &=&{{352\,{{\pi }^3}}\over {75}}  , \nn \\
  B_5 &=&{{11008\,{{\pi }^3}}\over {1485}} , \nn \\
  B_6 &=&{{5982208\,{{\pi }^3}}\over {2027025}} , \nn \\
  B_7 &=& {{-164096\,{{\pi }^3}}\over {57915}} , \nn \\
  B_8 &=& {{-952002848\,{{\pi }^3}}\over {172297125}} , \nn \\
  B_9 &=&{{-337870336\,{{\pi }^3}}\over {72747675}} , \nn \\
  B_{10} &=&{{-8650820224\,{{\pi }^3}}\over {4464061875}} . \nn
\eea
For $d=5$, $p=3$ and $p=2$:
\bea
  B_0 &=&10\,{{\pi }^3} , \nn \\
  B_1 &=&{{-80\,{{\pi }^3}}\over 3} , \nn \\
  B_2 &=&{{70\,{{\pi }^3}}\over 3} , \nn \\
  B_3 &=&{{-14\,{{\pi }^3}}\over 3} , \nn \\
  B_4 &=&{{-29\,{{\pi }^3}}\over {18}}  , \nn \\
  B_5 &=&{{-37\,{{\pi }^3}}\over {90}} , \nn \\
  B_6 &=&{{-{{\pi }^3}}\over {12}}
 , \nn \\
  B_7 &=&{{-53\,{{\pi }^3}}\over {3780}}
 , \nn \\
  B_8 &=&{{-61\,{{\pi }^3}}\over {30240}} , \nn \\
  B_9 &=&{{-23\,{{\pi }^3}}\over {90720}}
 , \nn \\
  B_{10} &=&{{-11\,{{\pi }^3}}\over {388800}}
 . \nn
\eea
For $d=5$, $p=1$:
\bea
  B_0 &=&5\,{{\pi }^3} , \nn \\
  B_1 &=&{{-10\,{{\pi }^3}}\over 3} , \nn \\
  B_2 &=&{{-10\,{{\pi }^3}}\over 3} , \nn \\
  B_3 &=&{{2\,{{\pi }^3}}\over 3} , \nn \\
  B_4 &=& {{35\,{{\pi }^3}}\over {18}} , \nn \\
  B_5 &=&{{91\,{{\pi }^3}}\over {90}} , \nn \\
  B_6 &=&{{-{{\pi }^3}}\over {12}} , \nn \\
  B_7 &=&{{-2101\,{{\pi }^3}}\over {3780}} , \nn \\
  B_8 &=&{{-3289\,{{\pi }^3}}\over {6048}} , \nn \\
  B_9 &=&{{-32791\,{{\pi }^3}}\over {90720}} , \nn \\
  B_{10} &=&{{-104873\,{{\pi }^3}}\over {544320}} . \nn
\eea
For $d=4$, $p=2$:
\bea
  B_0 &=&16\,{{\pi }^2} , \nn \\
  B_1 &=&-32\,{{\pi }^2} , \nn \\
  B_2 &=&{{304\,{{\pi }^2}}\over {15}} , \nn \\
  B_3 &=&{{-160\,{{\pi }^2}}\over {63}} , \nn \\
  B_4 &=& {{-176\,{{\pi }^2}}\over {315}} , \nn \\
  B_5 &=&{{-608\,{{\pi }^2}}\over {3465}} , \nn \\
  B_6 &=&{{-11104\,{{\pi }^2}}\over {135135}} , \nn \\
  B_7 &=&{{-448\,{{\pi }^2}}\over {8775}} , \nn \\
  B_8 &=&{{-443504\,{{\pi }^2}}\over {11486475}} , \nn \\
  B_9 &=&{{-467045792\,{{\pi }^2}}\over {13749310575}} , \nn \\
  B_{10} &=&{{-2327539744\,{{\pi }^2}}\over {68746552875}} . \nn
\eea
For $d=4$, $p=1$:
\bea
  B_0 &=&{{32\,{{\pi }^2}}\over 3} , \nn \\
  B_1 &=&{{-32\,{{\pi }^2}}\over 3} , \nn \\
  B_2 &=& {{-32\,{{\pi }^2}}\over {45}} , \nn \\
  B_3 &=&{{352\,{{\pi }^2}}\over {189}} , \nn \\
  B_4 &=& {{928\,{{\pi }^2}}\over {945}} , \nn \\
  B_5 &=& {{1952\,{{\pi }^2}}\over {10395}} , \nn \\
  B_6 &=&{{-38848\,{{\pi }^2}}\over {405405}} , \nn \\
  B_7 &=&{{-262336\,{{\pi }^2}}\over {2027025}} , \nn \\
  B_8 &=&{{-3454688\,{{\pi }^2}}\over {34459425}} , \nn \\
  B_9 &=&{{-3024736672\,{{\pi }^2}}\over {41247931725}} , \nn \\
  B_{10} &=&{{-12244948288\,{{\pi }^2}}\over {206239658625}} . \nn
\eea
For $d=3$, $p=2$ and $p=1$:
\bea
  B_0 &=&6\,{{\pi }^2} , \nn \\
  B_1 &=&-6\,{{\pi }^2} , \nn \\
  B_2 &=& {{\pi }^2} , \nn \\
  B_3 &=&{{{{\pi }^2}}\over 3} , \nn \\
  B_4 &=&{{{{\pi }^2}}\over {12}}  , \nn \\
  B_5 &=&{{{{\pi }^2}}\over {60}} , \nn \\
  B_6 &=&{{{{\pi }^2}}\over {360}} , \nn \\
  B_7 &=&{{{{\pi }^2}}\over {2520}} , \nn \\
  B_8 &=&{{{{\pi }^2}}\over {20160}} , \nn \\
  B_9 &=&{{{{\pi }^2}}\over {181440}} , \nn \\
  B_{10} &=&{{{{\pi }^2}}\over {1814400}} . \nn
\eea
And for $d=2$, $p=1$:
\bea
  B_0 &=&8\,\pi  , \nn \\
  B_1 &=&{{-16\,\pi }\over 3} , \nn \\
  B_2 &=&{{8\,\pi }\over {15}} , \nn \\
  B_3 &=&{{32\,\pi }\over {315}} , \nn \\
  B_4 &=&{{8\,\pi }\over {315}}  , \nn \\
  B_5 &=&{{32\,\pi }\over {3465}} , \nn \\
  B_6 &=&{{3056\,\pi }\over {675675}} , \nn \\
  B_7 &=&{{1856\,\pi }\over {675675}} , \nn \\
  B_8 &=&{{22664\,\pi }\over {11486475}} , \nn \\
  B_9 &=&{{4481632\,\pi }\over {2749862115}} , \nn \\
  B_{10} &=&{{104409808\,\pi }\over {68746552875}} . \nn
\eea

\bce {\large \bf V. Transversal Laplacian and non-local counterterms} \ece

In this section we demonstrate that the heat kernel expansion
for the Hodge-de Rham Laplacian on transversal $p$-forms
contains a constant term even in odd--dimensional spaces.
Let us introduce the following notation:
\beq
\kappa (t)=\sum_{k=1}^\infty \exp (-tk^2).
\label{eq:kappa}
\eeq
The asymptotic behavior of $\kappa$ in the limit $t\to 0$ is well
known
\beq
\kappa (t)= \frac 12 \left ( \sqrt{\frac \pi{t}} -1 \right ) ,
\label{eq:kas}
\eeq
where corrections are exponentially small and can be neglected
in the computation of power-low asymptotics.

The transversal heat kernel
\beq
K(t;p,d)=\sum_{l=1}^\infty D_l(p,d) \exp \left[tb_l(p,d)\right]
\label{eq:kpd}
\eeq
on odd-dimensional spheres
can be expressed in terms of $\kappa$ using the explicit form
of $D_l$ and $b_l$ of Sect. 2:
\begin{eqnarray}
K(t;0,3)&=&-1-\kappa '(t), \nonumber \\
K(t;1,3)&=&-2\left[\kappa '(t)+\kappa (t)\right], \nonumber \\
K(t;0,5)&=&-1+\frac 1{12}e^{4t} \left[
\kappa ''(t)+\kappa '(t)\right], \nonumber \\
K(t;1,5)&=&1+\frac 13 e^t \left[\kappa ''(t) +4\kappa '(t)\right],
 \nonumber \\
K(t;2,5)&=&\frac 12 \left[
\kappa ''(t)+5\kappa '(t)+4\kappa (t)\right], \nonumber \\
K(t;0,7)&=&-1-\frac {e^{9t}}{360}\left[\kappa '''(t)+5\kappa ''(t)
+4\kappa '(t) \right], \nonumber \\
K(t;1,7)&=&1-\frac {e^{4t}}{60} \left[\kappa '''(t)+10\kappa ''(t)
+9\kappa '(t) \right], \nonumber \\
K(t;2,7)&=&-1-\frac {e^t}{24} \left[\kappa '''(t)+13\kappa ''(t)
+36 \kappa '(t) \right], \nonumber \\
K(t;3,7)&=&-\frac 1{18} \left[\kappa '''(t)+14 \kappa ''(t)
+41 \kappa '(t)+36 \kappa (t)\right], \label{eq:kkpd}
\end{eqnarray}
where the prime denotes differentiation with respect to the argument.

Now, we can evaluate the coefficient $a_{d/2}$ before $t^0$ in
the small-$t$ expansion of $K(t;p,d)$. Derivatives of $\kappa$
do not contribute to this coefficient. One obtains the following
remarkable relation:
\beq
a_{d/2}=(-1)^{p+1} . \label{eq:ad2}
\eeq
At first sight this relation contradicts the general theory \cite{Gil}
of the heat kernel, which precludes integer powers of $t$ on
odd-dimensional manifolds without boundary. In fact, the relation
(\ref{eq:ad2}) can be derived from general formulae \cite{Gil}.
Consider an odd-dimensional manifold $M$ without boundary. The space of
$p$-forms $\Lambda^p$ can be decomposed in a direct sum of eigenspaces
of the Hodge-de Rham Laplacian:
\beq
\Lambda^p=\Lambda^{pT}\oplus \Lambda^{pL} \oplus H^p ,
\label{eq:lamp}
\eeq
where $\Lambda^T$ and $\Lambda^L$ are transversal and longitudinal
$p$-forms respectively. $H^p$ denotes the space of harmonic $p$-forms
spanned by zero modes of the Hodge-de Rham Laplacian. The Laplace
operator on {\it all \/} $p$-forms satisfies the whole set of requirements
in \cite{Gil} and, hence, the corresponding coefficient in front of
$t^0$ in the heat kernel expansion should vanish. On the other hand,
this coefficient is just the sum of the coefficients in front of
$t^0$ for the same operator restricted to the spaces on the
right hand side of (\ref{eq:lamp}). But this immediately
gives:
\beq
0=a_{d/2}^p+a_{d/2}^{p-1}+\beta_p , \label{eq:recur}
\eeq
where, as above, $a_{d/2}$ denotes the constant term in the heat
kernel expansion for transversal $p$-forms. Here $\beta_p=$ dim $H^p$ is
the Betti number. In particular, for $0$-forms we have
\beq
a_{d/2}^0=-\beta_0 .\label{eq:ap0}
\eeq
The two equations (\ref{eq:recur}) and (\ref{eq:ap0}) can be solved
giving
\beq
a_{d/2}^p=\sum_{q=0}^p (-1)^{p-q+1} \beta_q . \label{eq:ad2g}
\eeq
The relation (\ref{eq:ad2}) is a particular case of (\ref{eq:ad2g}).

Consider now the quantum path integral for an antisymmetric tensor
field with the action (\ref{eq:act}). The partition function $Z_p$
can be expressed in terms of the determinants of the Hodge-de Rham
Laplacian on transversal forms \cite{qten1,qten2}
\beq
Z_p=\prod_{q=0}^p \det_{qT} (-\Delta_{\rm HdR})^{-\frac 12 (-1)^{p-q}}.
\label{eq:zp}
\eeq
To avoid possible ambiguities in treating the zero modes we suppose
that $\beta_p=0$. The ``total" heat kernel for $Z_p$ is just an
alternated sum of heat kernels for transversal forms. Since the
coefficient multiplying $t^0$ leads to a logarithmic divergence in
the path integral, on an odd-dimensional manifold without
boundary we have that this divergence is proportional to
\beq
\sum_{q=0}^p (p-q+1)(-1)^{p-q+1} \beta_q .
\label{eq:div}
\eeq
Such divergency cannot be cancelled by means of an integral of a local
invariant constructed from the Riemann tensor and, hence, it requires
a non--local counterterm.

Some topological effects in quantum theories of antisymmetric
tensor field were discussed in \cite{qten2}. These effects
are however related to the Gauss--Bonnet term, which can be
expressed in function of local densities and vanishes for odd--dimensional
spheres.


\newpage

\noindent{\bf Acknowledgments}

We thank Klaus Kirsten for very interesting comments.
ML thanks all the members of the Department ECM, Barcelona University,
for warm hospitality.
This work was initiated during a stay of EE at
the Institute of Theoretical Physics,
Chalmers University of Technology (Sweden). 
The stay of ML in Barcelona was made possible by the ERASMUS program.
EE was supported by DGICYT and  the Ministry of Foreign Affairs
(Spain). DV was supported by the Russian
Foundation for Fundamental Research and by GRACENAS, project
M94-2.2$\Pi$-18.


\newpage


\begin{thebibliography}{99}
\bibitem{CanW}
P. Candelas and S. Weinberg, Nucl. Phys. {\bf B237} (1984) 397.
\bibitem{FrT}
E.S. Fradkin and A.A. Tseytlin, Nucl. Phys. {\bf B227} (1983) 252.
\bibitem{PBG}
P.B. Gilkey, {\it Invariance theory, the heat equation and the
Atiyah--Singer theorem} (Publish or Perish, Delaware, 1984)
\bibitem{RO}
M.A. Rubin and C. Ordonez, J. Math. Phys. {\bf 26} (1985) 65.
\bibitem{IK} I. Iwasaki and K. Katase, Proc. Japan Acad. Sci. {\bf 55A}
(1979) 141.
 \bibitem{LSV}
V.D. Lyakhovsky, N.N. Shtykov and D.V. Vassilevich,
Lett. Math. Phys. {\bf 21} (1991) 89;
D.V. Vassilevich and N.N.Shtykov, Yadern. Fiz. {\bf 53} (1991) 869;
Teor. Mat. Fiz. {\bf 90} (1992) 12;
D.V. Vassilevich, Lett. Math. Phys. {\bf 26} (1992) 147.
\bibitem{Dplb}
D.V. Vassilevich, Phys. Lett. {\bf B348} (1995) 39.
\bibitem{SaSt}
A. Salam and J. Strathdee, Ann. Phys. {\bf 141} (1982) 556.
\bibitem{BaRon}
A. Barut and R. Raczka, {\it Theory of group representations and
applications} (PWN, Warszawa, 1977).
\bibitem{Djmp}
D.V. Vassilevich, J. Math. Phys. {\bf 36} (1995) 3174.
\bibitem{InPr}
E. Elizalde, M. Lygren and D. Vassilevich, in preparation.
\bibitem{BeMi}
B.L. Beers and R.S. Millman, Amer. J. Math. {\bf 99}
(1977) 801.
\bibitem{IkTa}
A. Ikeda and Y. Taniguchi, Osaka J. Math. {\bf 15} (1978) 515.
\bibitem{eet1}
  E. Elizalde, {\sl On two complementary approaches aiming at the
definition of the determinant of an elliptic partial differential
operator}, Trento preprint UTF 359, hep-th/9508167 (1995).

\bibitem{eli1} E. Elizalde, J. Math. Phys. {\bf 35}
(1994) 3308.

\bibitem{eli2} E. Elizalde, J. Math. Phys. {\bf 35}
(1994) 6100.

\bibitem{eli3}
E. Elizalde, {\it Ten physical applications of spectral zeta functions}
(Springer-Verlag, Berlin, 1995);
E. Elizalde, S.D. Odintsov, A.
Romeo, A.A. Bytsenko and S. Zerbini, {\it Zeta regularization
techniques with applications} (World Sci., Singapore, 1994).

\bibitem{eemc} E. Elizalde, Math. Computation {\bf 47} (1986) 347.
\bibitem{bek} M. Bordag, E. Elizalde and K. Kirsten, {\sl
Heat-kernel coefficients of the Laplace operator on the $D$-dimensional
ball},  J. Math. Phys., to appear.

\bibitem{bgke} M. Bordag,  B. Geyer,  K. Kirsten and E. Elizalde, {\sl
Zeta-function determinant of the Laplace operator on the
$D$-dimensional ball},
Commun. Math. Phys., to appear.

\bibitem{Gil}
P.B. Gilkey, J. Diff. Geom. {\bf 10} (1975) 601.
\bibitem{qten1}
W. Siegel, Phys. Lett. {\bf 93B} (1980) 170;
E. Sezgin and P. van Nieuwenhuizen, Phys. Rev. {\bf D32} (1980) 301;
T.Kimura, Prog. Theor. Phys. {\bf 65} (1981) 338;
I.A. Batalin and G.A.Vilkovisky, Phys. Lett. {\bf 120B} (1983) 166.
\bibitem{qten2}
M.J. Duff and P. van Nieuwenhuizen, Phys. Lett. {\bf 94B} (1980) 179;
Y.N. Obukhov, Phys. Lett. {\bf 109B} (1982) 195.

\end{thebibliography}
\end{document}